# Nanoscale Superconducting Honeycomb Charge Order in IrTe$_2$

*Hyo Sung Kim[1,2], Sooran Kim[2], Kyoo Kim[2], Byung Il Min[2], Yong-Heum Cho[2,3], Lihai Wang[2,3], Sang-Wook Cheong[2,3,4], Han Woong Yeom[1,2]\**

[1]Center for Artificial Low Dimensional Electronic Systems, Institute for Basic Science (IBS), Pohang 790-784, Korea

[2]Department of Physics, Pohang University of Science and Technology, Pohang 790-784, Korea

[3]Laboratory for Pohang Emergent Materials, Pohang University of Science and Technology, Pohang 790-784, Korea

[4]Rutgers Center for Emergent Materials and Department of Physics and Astronomy, Piscataway, New Jersey 08854, USA

\*To whom all correspondence should be addressed; yeom@postech.ac.kr. Fax) +82542799889.

Abstract

   Entanglement of charge orderings and other electronic orders such as superconductivity is in the core of challenging physics issues of complex materials including high temperature superconductivity. Here, we report on the observation of a unique nanometer scale honeycomb charge ordering of the cleaved $IrTe_2$ surface, which hosts a superconducting state. $IrTe_2$ was recently established to exhibit an intriguing cascade of stripe charge orders. The stripe phases coexist with a hexagonal phase, which is formed locally and falls into a superconducting state below 3 K. The atomic and electronic structures of the honeycomb and hexagon pattern of this phase are consistent with the charge order nature but the superconductivity does not survive on neighboring stripe charge order domains. The present work provides an intriguing physics issue and a new direction of functionalization for two dimensional materials.

Keywords: $IrTe_2$, Transition metal dichalogenides, Charge order, Stripe charge order, Honeycomb charge order, Superconductivity

Charge orderings due to Coulomb interaction and charge density waves (CDW) due to electron-phonon interaction have been among the key physics issues in understanding complex electronic systems with entangled quantum states. The suppression of CDW order leads to the emerging superconductivity[1,2], the CDW coexists with high temperature superconductivity in cuprates[3,4], and the charge orderings are entangled with colossal magnetoresistance in manganese oxides[5,6]. In recent years, the charge ordering and CDW phenomena coupled with superconductivity are attracting renewed interest for the new functionality of two dimensional transition metal dichalogenides[7-10]. The recent development of microscopy techniques with the spectroscopic capability such as scanning tunneling microscopy/spectroscopy (STM/S) has made it possible to disclose atomic scale details of charge orderings in complex low dimensional systems in coexistence and competition with various electronic orders. Most of long range charge orderings found so far in such systems are stripe structures[11] while hexagonal structures are common for CDW systems[12–14]. In addition, microscopic studies identified the checkerboard-type charge orderings in a high temperature superconductor[15] and a manganite perovskite[16,17]. The origin of the checkerboard order is still not fully understood but it corresponds to the intrinsic or extrinsic overlap of two stripe orders[18].

In this study, we identify with STM a unprecedented honeycomb charge ordering in nanoscale coexistence with stripe charge orders of $IrTe_2$[19–24], which is distinct from the hexagonal CDW. Moreover, this honeycomb charge order hosts superconductivity below 3K in clear distinction from the coexisting stripe orders. That is, in this material, the superconductivity is uniquely and intriguingly coupled to a specific type of charge order. $IrTe_2$ is currently under extensive investigations due to its unusual electronic properties related to the charge ordering. $IrTe_2$ is an outstanding transition metal dichalcogenide with a charge order (Ir $5d^{3+}$- Ir $5d^{4+}$) transition into stripe phases with a large hysteresis and a nontrivial remnant metallic conductivity[19–23,25,26]. Upon doping, the stripe charge ordering is suppressed and the superconductivity emerges, suggesting a quantum critical behavior[27–31]. The driving force of the transition into the stripe orders is not fully understood yet but the CDW nature was ruled out[32–34]. The interplay of various different degrees of freedom is widely recognized, such as the strong structural distortions involving the Ir-Ir dimerization and the depolymerization of the Te-Te interlayer bonding, the Ir 5$d$ charge disproportionation, the strong reorganization of Fermi surfaces, and the substantial spin-orbit coupling. While recent findings seem to converge into the picture of a cooperative Jahn-Teller transition through Ir dimerization, the role of other degrees of freedom is not understood[35].

The stripe orders here are shown to coexist with a distinct hexagonal phase. Figure 1a shows a cleaved IrTe$_2$ surface where a Te layer is exposed[36–38]. As reported previously, various stripe phases with three different orientations (arrows in the figure) and with different spacings of ×3 (tripled periodicity), ×5, ×8, and ×11 are observed as inhomogeneous mixture (Fig. 1e)[21–23]. In extra, we find another distinct phase with a hexagonal symmetry as enclosed by stripe phases of three different orientations on the surface below the transition temperature. Figure 1c shows one such domain, which is composed of hexagons and a honeycomb wall lattice of between 5 (1.75 nm) and 8 times (2.8 nm) the lattice constant[36,37]. The Fourier transformation of the STM image (Fig. 1b) indicates a quasi long-range hexagonal order with an average periodicity of 2.2 nm. Note also that the surface interatomic distance is carefully calibrated to be, indicating a substantial (10~11 %) strain on the surface layer, which is reflected in the large scale corrugation of the cleaved surfaces (Fig. S1 in Supporting information). The areal ratio between the stripe and hexagonal phases varies on different samples, cleavages, and parts of the surface but we always find nonnegligible portion of hexagonal domains. A large domain of it can extend to a few hundred nm, which has a tendency to be enlarged after cooling samples rapidly (see Fig. 1d and Fig. S1 in Supporting information).

Figures 1e–f compare detailed atomic scale structures of the stripe and the hexagonal phase. The stripe phase exhibits rows of atoms with the periodically modulated contrast, which was well characterized as due mainly to the vertical structural modulation of the Te surface layer[23]; the Te rows with the bright contrast are buckled up due to the dimerization of Ir rows underneath[19,32,33]. There exist three dark rows and two bright ones in the most widely found stripe phase of a ×5 periodicity, underneath which Ir atomic rows of $5d^{3+}$ and $5d^{4+}$ states are located, respectively[23]. In contrast, the hexagonal phase shows quasi regular hexagons of the bright contrast as surround by a honeycomb wall lattice of the dark contrast. We note that the dark and bright contrast in the STM topography (the apparent height in the corresponding line profiles of Figs. 1e–f) are consistent in both phases in the whole tunneling bias range (see Fig. S2 in Supporting information). That is, it is rather straightforward that the hexagonal structure consists of similar structural motifs, the Te buckling and Ir dimers, to the stripe phase.

The similarity of the stripe and the hexagonal phase is not limited to their atomic structures. Figure 2a shows STS spectra taken for bright and dark atomic rows of the stripe and the hexagonal phase. These spectra are almost identical, indicating that electronic structures of two phases are almost the same. The prominent peak and dip structures in the energy range shown is thought to be due to the Te depolymerization or buckling, which is intimately entangled with the Ir charge ordering and dimerization[19,32,33,38]. The consistent electronic structure of the stripe and the hexagonal phase can further be confirmed by $dI/dV$ maps (Figs. 2b–q), which reveal the atomic scale lateral variation of local density of states (LDOS) in the surface Te layer. As detailed previously[23], in the stripe phase, the LDOS at +0.3, -0.5 and -0.9 (-0.3, +0.5 and +0.9) eV is enhanced at bright (dark) atomic rows of the topography indicating directly the electronic modulation in the Te layer. This LDOS modulation is copied in the hexagonal phase although there is some extra symmetry breaking (Figs. 2h and 2q). In short, LDOS maxima are located along the dark trenches (hexagon boundaries) in 0.3–1.0 eV but move towards the bright ridges (into hexagon centers) in the stripe (hexagonal) phase approaching the Fermi level in empty states. On the other hand, in filled states, LDOS maxima sit on the dark trenches in the low bias but move towards the ridge (hexagon centers) at a higher bias such as -0.9 eV. In the transition region, at -0.5 and +0.3 eV, the maxima split for stripes to make the maps more complicated for hexagons. We thus can conclude that the stripe and the hexagonal phase have very similar atomic and electronic structures, which are based on the dimerization and charge ordering of Ir atoms. This means that the bright hexagon in the STM image of the hexagonal phase is composed of protruded Te atoms with the dimerized Ir $5d^{4+}$ atoms underneath and the dark honeycomb walls correspond to Te atoms bonded to Ir $5d^{3+}$ atoms.

The present hexagonal phase of a three-fold symmetry and a honeycomb charge distribution is unique among charge order materials, while it can be compared with the checkerboard charge ordering for two fold symmetric crystals[15,18]. In the case of CDW, domain walls of incommensurate orderings widely exhibit the competition between stripe and hexagonal phases[12–14,39] as is general for incommensurate superstructures in 2D. The uniqueness of the present system is that the present stripe orders are commensurate ones and is not a CDW system and not even insulating. The similarity between the present system and 2D incommensurate phases, but, suggests that the honeycomb phase can be a 2D ordering in contrast to the nontrivial interlayer coupling of the stripe phase[40]. In surface layers, a different ordering is likely through the lack of the Te-Te interlayer bonding towards the missing top layer and the possible surface strain. The surface strain is as large as 11% in the present case (Fig. 1d). However, the very recent STM study disclosed a disordered hexagonal phases very similar to the present one on doped samples with the superconducting ground state[20]. We also confirmed this with our own doped samples (Fig. S3 in Supporting information). Moreover, as mentioned above, the hexagonal phase is enhanced for the rapid thermal quenching, where the bulk stripe ordering is partly prohibited too[41]. The occurrence of the hexagonal phase on doped and quenched samples indicates that it is not a simple surface effect. We suggest that the hexagonal phase has a 2D nature, the reduced interlayer coupling, which can be induced by dopants, disorders, and local strains. This also indicates that the hexagonal order is not formed uniformly over the bulk material as the previous x-ray diffraction study did not detect it[26, 38].

Based on the above discussion, we try to construct a 2D structure model of the hexagonal phase from that of the stripe phase. As shown in Figs. 2b and 2j, the stripe phase is composed of three dark contrast rows (blue atoms) and two bright ones (red), under which $Ir^{3+}$ and $Ir^{4+}$ rows exist, respectively[19,32,33]. Since there exist three rotationally degenerate ×5 stripes, we can simply overlap these three structures. This idea is consistent with the fact that the hexagonal phase domain is in most cases enclosed with stripe phases with three different orientations. This overlap obviously yields the hexagonal unit cell with the dark trenches as the honeycomb walls. Within this model, one can understand the varying size of hexagons based on the coexistence and competition of the ×3, ×5 and ×8 stripe orders.

The feasibility of the present model is checked by first principles density-functional theory (DFT) calculations using a relatively small unit cell of 6×6×1. We found that the hexagonal structure is not stabilized in the bulk configuration but can be stabilized for the monolayer under the compressive strain of about 10 %, which is consistent with the STM measurement. The relaxed structure is shown in Fig. 3b, where a hexagon is composed of twenty-seven Ir atoms and twelve of them form Ir tetramers with strong Ir-Ir bonding (yellow rods). As suggested above, hexagons and the honeycomb wall network are mainly composed of $Ir^{4+}$ and $Ir^{3+}$ (or less charged than 4+) atoms, respectively. This corresponds to one of the smallest hexagons observed. The height distribution of Te surface atoms in this structure is shown in Fig. 3a, which is qualitatively consistent with the STM topography of Figs. 2f and 2n. Further refinements of the model is obviously required since the LDOS maps of Fig. 2h and 2q indicate a much higher degree of the honeycomb symmetry than the present calculation and an extra symmetry breaking of the hexagons. However, we suggest that this model is a reasonable starting point, which strongly indicates the charge order, 2D, and strain origin of the honeycomb phase.

Most interestingly, the superconductivity is directly related with the hexagonal phase. Figure 4b shows a *dI/dV* line scan obtained at 1.05 K across a domain boundary between the hexagonal and stripe phase (white dashed line in Fig. 4a). A clear superconducting gap with a size of 0.6 meV is formed on the hexagonal domain (Fig. 4b), which decays into the stripe domain. The typical exponential decay of the gap is quantified to yield a superconducting coherence length of ~50 nm. A similar value, ~55 nm, is obtained from the measurement using a vortex induced by the magnetic field (Fig. 4d). This confirms clearly that the stripe phase and the superconductivity is mutually exclusive. As shown in Figs. 4e and 4g, the superconducting gap disappears above ~3 K and the transition temperature is rather accurately estimated as 3.1 K. Under a magnetic field (Fig. 4d and Fig. 4f), the gap starts to break from 0.07 T, corresponding to the lower critical field ($H_{c1}$), and is totally broken at 0.1 T ($H_{c2}$). These results solidly corroborate the superconductivity origin of the gap. The occurrence of an irregular hexagonal pattern on the surface of doped samples with superconductivity at low temperature can also be understood consistently[27–31]. As mentioned above, the rapid quenching of the sample prohibits the long range ordering of the stripe phase and the enhanced hexagonal domains. This change was confirmed by bulk structural and electric measurements while the bulk superconductivity of a quenched sample without a doping was not observed yet.

The essence of the present model, the microscopic overlap of three rotated charge order $q$'s, is similar to the $3q$ model of the skyrmion in the 2D triangular spin system[42]. The stabilization of the $3q$ state is basically a nonlinear effect due to the higher order term in the Landau-Ginzburg Free energy. Consistent to the present picture, the local-strain-induced stripe phase was recently reported for $NbSe_2$ with the host hexagonal CDW ordering[43]. Combined with the present finding, this indicates that the competition between $1q$ and $3q$ state is general, irrespective of the detailed mechanism and energetics of charge orderings.

The relationship between the charge order and the superconductivity is not a simple mutual exclusion as also noticed very recently for high temperature superconductivity[44]. Moreover, this relationship cannot be direct since the energy scale of the charge order is huge with the Ir dimerization energy of the order of one eV. One can instead note the reduced interlayer coupling and the strain in the hexagonal one, which can be important for the superconductivity. The 2D hexagonal Fermi surface made through the reduced interlayer coupling would have definite merit for the Cooper pairing over those of the stripe phase with a strong 3D modulation. On the other hand, the nematic electronic fluctuation would be frozen into the static stripe order lattice while it can still survive in the hexagonal phase to enhance the Cooper-pair instability[45]. While the coexistence of this particular type of charge orders and the superconductivity is not fully understood at present, the nanoscale coexistence of unusual superconductivity with the charge-ordered metallic state provides a unique lateral heterointerface for further exploitation of the functionality, such as quantum thermal and interference devices using proximity junctions[46-48], of two dimensional material.

AUTHOR CONTRIBUTION

Hyo Sung Kim carried out the STM measurements. Soo-Ran Kim, Kyoo Kim and Byung Il Min performed the DFT calculations. Yong-Heum Cho, Lihai Wang and Sang-Wook Cheong synthesized the crystals. Hyo Sung Kim and Han Woong Yeom analysed the data and wrote the manuscript. Han Woong Yeom supervised the research.

ACKNOWLEDGMENT

This work is supported by Institute for Basic Science (Grant No. IBS-R014-D1). YHC and SWC are partially supported by the Max Planck POSTECH/KOREA Research Initiative Program (Grant No. 2011-0031558) through NRF of Korea funded by MEST. The work at Rutgers was funded by the Gordon and Betty Moore Foundation's EPiQS Initiative through Grant GBMF4413 to the Rutgers Center for Emergent Materials.We acknowledge Tae-Hwan Kim for the technical help of in STM measurements and Ki-Seok Kim, Illya Eremin, and Alireza Akbari for theoretical discussion.

SUPPLEMENTARY MATERIALS

Details of the honeycomb phase depending on the cooling process are provided. In addition supporting bias dependent STM topographies are available.

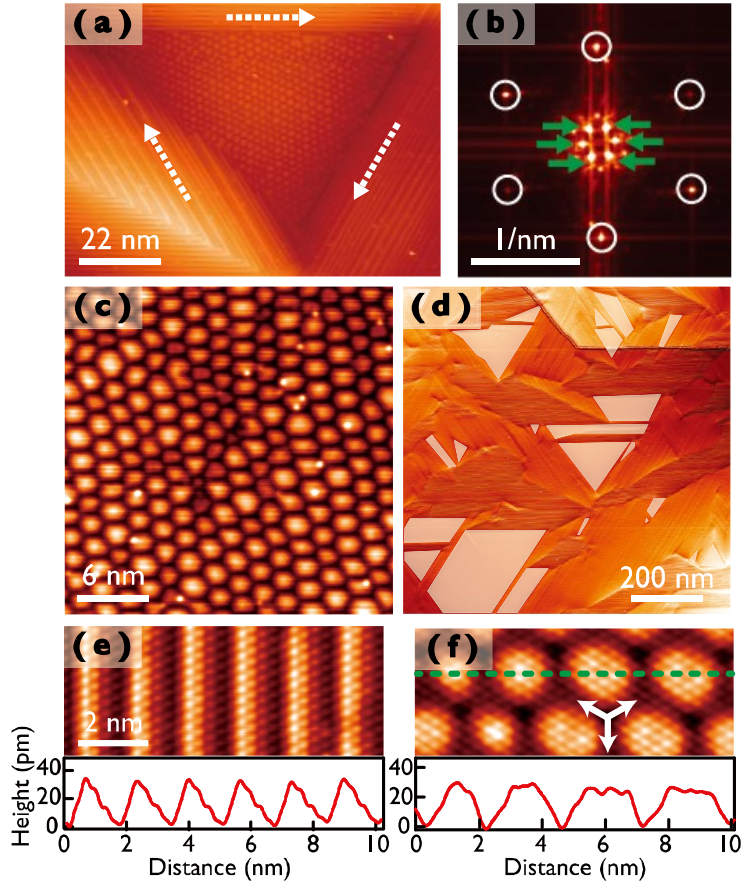

**Figure 1.** STM topographies of the cleaved IrTe$_2$ at 78 K. **(a)** A typical cleaved IrTe$_2$ surface shows hexagonal and stripe phases coexisting. **(b)** The FFT image of the hexagonal phase domain shown in **(c)**. White circles and green arrows indicate the periodicities of the atoms and hexagons (0.35 nm and 2.2 nm, respectively). **(d)** Large area image where the hexagonal (white shaded) and stripe phase coexist. **(e)** Atom-resolved STM topography of the ×5 stripe phase, and **(f)** that of the hexagonal structure. Tunneling biases are V$_s$ = 1 V for **(a)**, **(c)** and **(d)** and 20 mV for the atom-resolved images of **(e)** and **(f)**. STM line profiles crossing the stripes and hexagons [along the dashed line in **(f)**] are also given for **(e)** and **(f)**. The white arrows in **(f)** indicate the crystalline direction of IrTe$_2$.

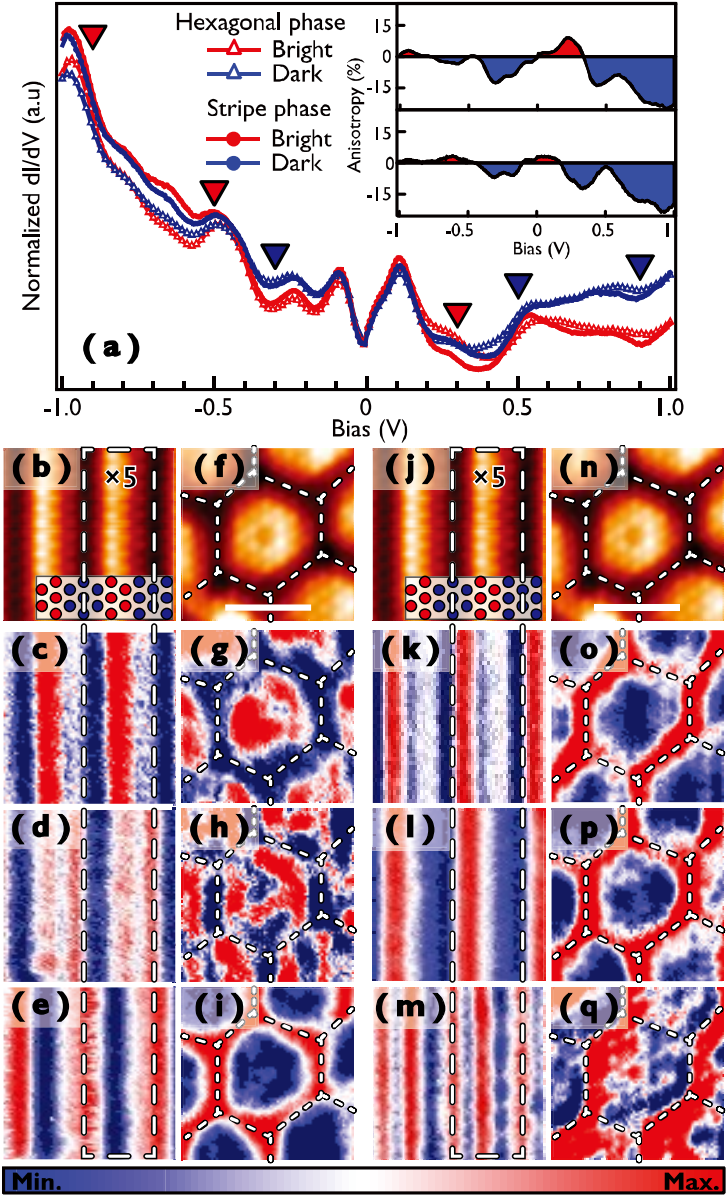

**Figure 2.** *dI/dV* maps and STS spectra at 78K. **(a)** Averaged STS (*dI/dV*) spectra from bright and dark regions of both the hexagonal and the stripe phases. **(b)** and **(j)** A STM topography ($V_s$ = 20 mV) of the stripe phase and **(c)**, **(d)**, **(e)**, **(k)**, **(l)**, and **(m)** spatially resolved *dI/dV* maps [$V_s$ = -0.9, -0.5, -0.3, 0.9, 0.5, and 0.3 V, respectively, indicated by arrow heads in (a)] measured simultaneously. The corresponding data for the hexagonal phase are given in **(f)**, **(g)**, **(h)**, **(i)**, **(o)**, **(p)** and **(q)**. White dashed line indicates the single ×5 stripe and a hexagon unit and 2 nm scale bar is shown in **(f)** and **(n)**.

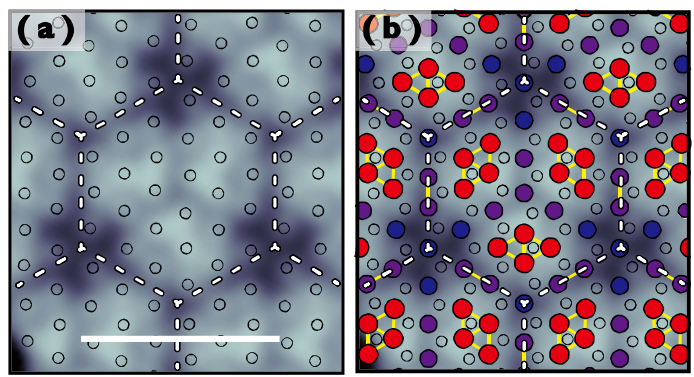

**Figure 3.** The DFT calculation results for monolayer IrTe$_2$. **(a)** The height profile of the Te surface atoms (circles) of the monolayer hexagonal structure shown in **(b)**, where the underlying Ir atoms of 4+ (red), 3+ (or less charged than 4+) (blue) states are also shown. The Ir-Ir bondings are connected by yellow rods. The 2 nm scale bar is shown in **(a)**.

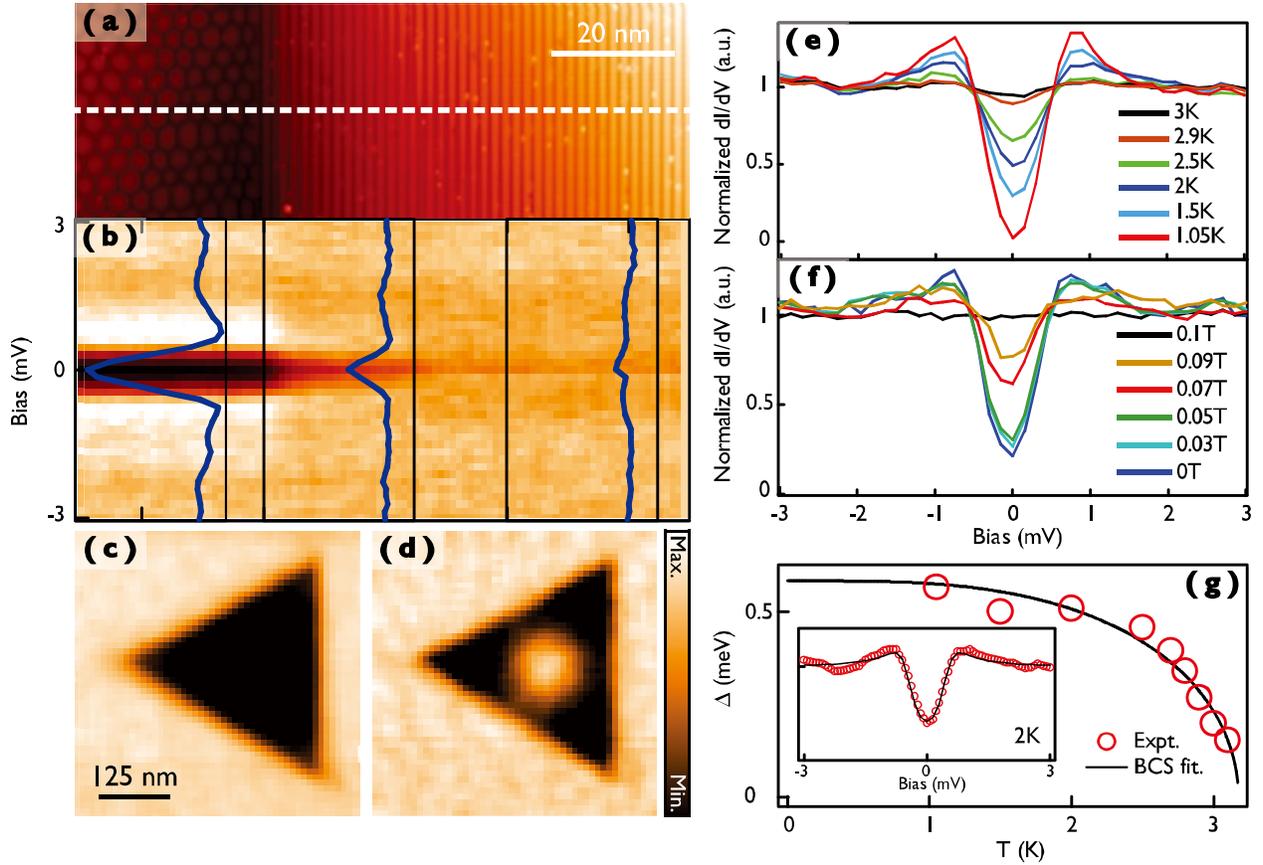

**Figure 4.** Proximity effects and superconducting state of the hexagonal phase at 1.1K. **(a)** Coexisting of the stripe and hexagonal phases. **(b)** Superconducting gap is vanished into the stripe phase. Three insets represent the characteristic gap of hexagonal, vicinity of hexagonal and stripe phase. **(c)** Zero bias conductance map of the hexagonal phase at 0 T. **(d)** Single vortex on the hexagonal phase at 0.07 T. Temperature dependence of the superconducting gap **(e)** and magnetic field dependence **(f)**. **(g)** BCS fitting for the experimental data. Inset shows a *dI/dV* curve of 2 K data and BCS theory.

For TOC only

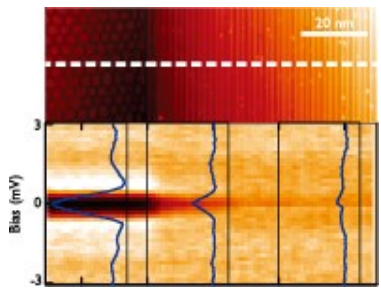